\documentclass{article}
\usepackage{spconf,amsmath,graphicx}

\usepackage{url}
\usepackage{threeparttable}
\usepackage{array}
\usepackage{multirow}
\newcolumntype{L}[1]{>{\let\newline\\\arraybackslash\hspace{0pt}}m{#1}}
\newcolumntype{C}[1]{>{\centering\let\newline\\\arraybackslash\hspace{0pt}}m{#1}}

\usepackage[pagebackref=true,breaklinks=true,hidelinks,colorlinks,bookmarks=false]{hyperref}

\renewcommand{\figurename}{Fig.}

\title{SRZoo: An Integrated Repository for Super-Resolution using Deep Learning}
\name{Jun-Ho Choi, Jun-Hyuk Kim, Jong-Seok Lee\thanks{
		This work was supported by the IITP grant funded by the Korea government (MSIT) (R7124-16-0004, Development of Intelligent Interaction Technology Based on Context Awareness and Human Intention Understanding). In addition, this work was also supported by Basic Science Research Program through the National Research Foundation of Korea (NRF) funded by the Korea government (MSIT) (NRF-2016R1E1A1A01943283).}}
\address{School of Integrated Technology, Yonsei University, Korea\\
	\{idearibosome, junhyuk.kim, jong-seok.lee\}@yonsei.ac.kr}

\begin{document}
%
\maketitle

\begin{abstract}
Deep learning-based image processing algorithms, including image super-resolution methods, have been proposed with significant improvement in performance in recent years.
However, their implementations and evaluations are dispersed in terms of various deep learning frameworks and various evaluation criteria.
In this paper, we propose an integrated repository for the super-resolution tasks, named SRZoo, to provide state-of-the-art super-resolution models in a single place.
Our repository offers not only converted versions of existing pre-trained models, but also documentation and toolkits for converting other models.
In addition, SRZoo provides platform-agnostic image reconstruction tools to obtain super-resolved images and evaluate the performance in place.
It also brings the opportunity of extension to advanced image-based researches and other image processing models.
The software, documentation, and pre-trained models are publicly available on GitHub\footnote{\url{https://github.com/idearibosome/srzoo}}.
\end{abstract}

\section{Introduction}
\label{sec:introduction}

In recent years, the performance of image processing algorithms has been significantly improved thanks to the development of deep learning-based approaches.
Massively enhanced computing environments power them in terms of both hardware (e.g., accelerated calculation by graphics processing units (GPUs)) and software (e.g., open-sourced versatile deep learning frameworks).
Many image-based tasks such as object classification, object detection, super-resolution, and image enhancement benefit from such improvements.

When introducing a new image processing algorithm, it is necessary to compare its performance with that of the other state-of-the-art algorithms to prove its improvement.
Various performance measures such as computation time and quantitative errors can be considered.
To make a fair comparison, it is necessary to run the algorithms in the same environment and use the same criteria.
However, we observe that such a fair comparison is not well employed for enhancement models such as super-resolution, unlike image classification and detection algorithms, because of the following reasons.

First, several integrated repositories provide state-of-the-art image classification and detection models, e.g., Keras Applications \cite{kerasapplications} and TensorFlow Detection Model Zoo \cite{tensorflowdetectionmodelzoo}.
However, there are very few repositories for the enhancement models, which makes employing such models harder than the classification models.
There exist some repositories for deep super-resolution models, e.g., VideoSuperResolution \cite{videosuperresolution} and super-resolution \cite{superresolution}.
However, they do not contain the official pre-trained models provided by the original authors.

\begin{figure}[t]
	\begin{center}
		\centering
		\includegraphics[width=1.0\linewidth]{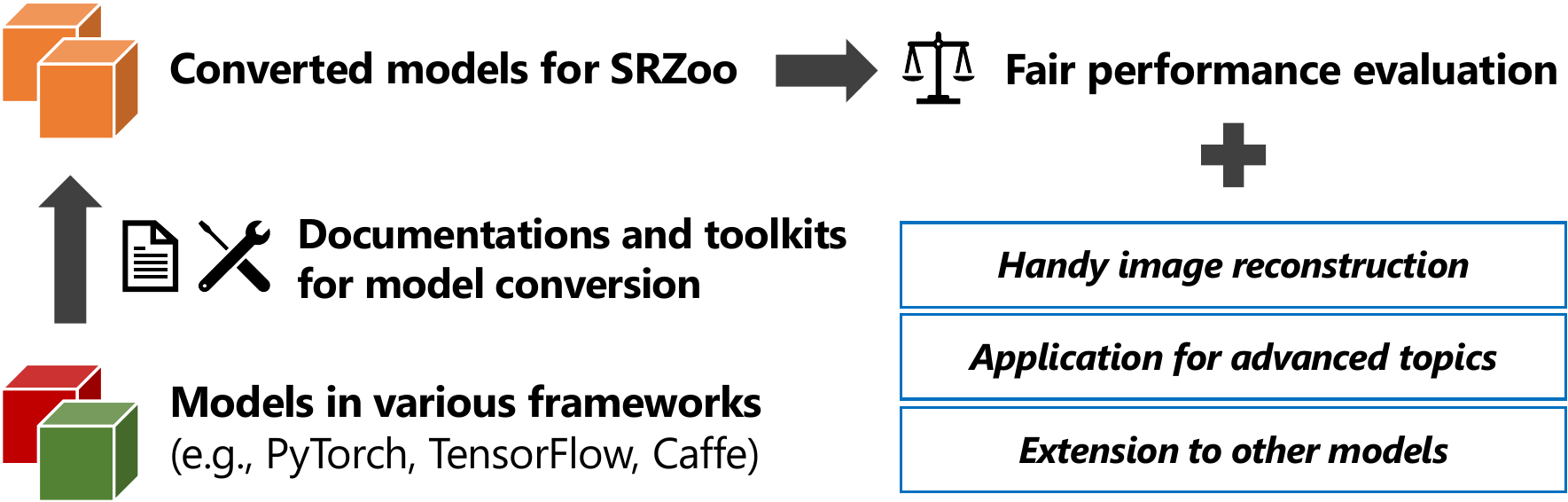}
	\end{center}
	\vspace{-6pt}
	\caption{Overall structure of SRZoo}
	\label{fig:structure}
\end{figure}

In addition, the performance evaluation metrics for the enhancement models are less standardized than those for the classification models.
In the classification tasks, the performance of different models has been justly compared using the same criteria, since the metrics are usually straightforward (e.g., counting the number of misclassified images).
However, in super-resolution tasks, we observe that the performance of the models is often evaluated using different criteria.
For example, different color spaces (e.g., RGB and YCbCr), different measurement implementations (e.g., different implementations of calculating structural similarity (SSIM)), and different post-processing methods (e.g., shaving edges and applying geometric self-ensemble \cite{lim2017enhanced}) are used.
Moreover, we find that most super-resolution methods are implemented on deep learning frameworks such as TensorFlow and PyTorch but are evaluated separately on MATLAB.

Motivated by these observations, we decide to develop an integrated repository for deep image super-resolution models, which provides various state-of-the-art pre-trained models that are ready to be deployed, and also additional useful tools.
Image super-resolution is one of the most evolutive research topics in recent days, which can be used in various applications such as surveillance, content streaming, and medical diagnosis \cite{yue2016image}.
Thus, our repository can catalyze to apply the state-of-the-art super-resolution models to such applications easily.
In addition, it also assists researchers to evaluate the performance with the same evaluation criteria.

\figurename~\ref{fig:structure} shows the overall structure of the proposed repository, which is named as SRZoo, the model zoo for the super-resolution models.
In this paper, we present the motivations, main features, performance comparison, and applications of our work in Sections~\ref{sec:motivations}, \ref{sec:features}, \ref{sec:performance}, and \ref{sec:applications}, respectively.

\section{Motivations}
\label{sec:motivations}

\subsection{Models with different frameworks}

The image super-resolution methods proposed in literature have been implemented in various deep learning frameworks such as Caffe \cite{hui2018fast}, PyTorch \cite{zhang2018image,wang2018esrgan,ahn2018fast}, and TensorFlow \cite{kim2018deep}.
Therefore, evaluating their performance requires setting up all of these frameworks in the same computing environment for a fair comparison.
In addition, when one needs to use the models to build an extension of the super-resolution task, it is necessary to implement it repeatedly for all the frameworks.
Examples of the extension include image classification using super-resolved images \cite{hao2018deep} and robustness analysis of super-resolution against adversarial perturbations \cite{choi2019evaluating}.

For the image classification and detection tasks, there exist integrated repositories that provide various state-of-the-art classification models on a single deep learning framework.
For example, Keras Applications \cite{kerasapplications} provides various deep image classification models with pre-trained weights.
It enables researchers to employ or modify the models for purposes rather than the original image classification task.

\subsection{Models with different evaluation criteria}
\label{sec:different_evaluation_criteria}

\begin{table}[]
	\footnotesize
	\setlength{\tabcolsep}{0.5em}
	\begin{center}
		\begin{tabular}{l | C{1.4cm} C{1.35cm} C{1.4cm} C{1.35cm}}
			\textbf{Method} & \textbf{Color channels} & \textbf{Shaving edges (pixels)} & \textbf{Image precision} & \textbf{Self-ensemble} \\
			\hline
			EDSR \cite{lim2017enhanced} & Y (YCbCr) & $\mathrm{scale}^{**}$ & Integer (8-bit) & Yes \\
			EDSR$^{*}$ \cite{lim2017enhanced} & RGB & $6+\mathrm{scale}^{**}$ & Integer (8-bit) & Yes \\
			RCAN \cite{zhang2018image} & Y (YCbCr) & $\mathrm{scale}^{**}$ & Float & Yes \\
			ESRGAN \cite{wang2018esrgan} & Y (YCbCr) & $\mathrm{scale}^{**}$ & Float & No \\
			RRDB \cite{wang2018esrgan} & Y (YCbCr) & $\mathrm{scale}^{**}$ & Float & No \\
			CARN \cite{ahn2018fast} & Y (YCbCr) & $\mathrm{scale}^{**}$ & Float & No
		\end{tabular}
	\end{center}
	\begin{tablenotes}
		\item \footnotesize{$^{*}$~On the DIV2K dataset}
		\item \footnotesize{$^{**}$~The number of pixels corresponding to the upscaling factor}
	\end{tablenotes}
	\caption{Conditions used for evaluating popular super-resolution methods.}
	\label{table:evaluation_criteria}
\end{table}

When developing a new algorithm, it is necessary to conduct performance evaluation under the same condition for both the new and existing algorithms.
However, we find that many super-resolution methods are evaluated with different criteria, which is often overlooked when the performance comparison is reported.
Table~\ref{table:evaluation_criteria} shows the evaluation methods used for some popular super-resolution models.
For instance, some super-resolution performance results are reported with the images converted to floating-point pixel values (e.g., using the ``$\mathrm{im2double}$'' function in MATLAB), whereas some other results are obtained from the original integer-point pixel values.
In addition, for calculating the SSIM values \cite{wang2004image}, some super-resolution methods employ the function included in MATLAB, while some others employ the code provided by the original authors, which produce slightly different results.

In the case of the super-resolution models, it is harder to standardize the evaluation criteria in comparison to the image classification models.
As the evaluation criterion of image classification, the success rate of classification such as top-$n$ error is usually used, which is straightforward and clearly defined.
However, there are multiple options to alter the evaluation procedure of super-resolution.
For example, there exist various configurations (e.g., color space, shaving edges) and evaluation methods (e.g., calculating pixel-wise differences, measuring structural similarity, considering perceptual awareness).
These factors make it difficult to compare different algorithms under the same condition.

\begin{table*}[t]
	\footnotesize
	\begin{center}
		\begin{tabular}{l | C{1.35cm} C{1.25cm} | C{1.35cm} C{1.25cm} C{1.25cm} C{1.7cm} C{1.7cm}}
			& \multicolumn{2}{c|}{\textbf{Reported}} & \multicolumn{5}{c}{\textbf{Measured}} \\
			\hline
			\textbf{Models (year)} & \textbf{PSNR (dB)} & \textbf{SSIM} & \textbf{PSNR (dB)} & \textbf{SSIM} & \textbf{NIQE} & \textbf{Running time (CPU, s)} & \textbf{Running time (GPU, s)} \\
			\hline
			ESRGAN (2018) \cite{wang2018esrgan} & N/A & N/A & 25.31 & 0.6502 & 3.664 & ~~8.236 & 0.106 \\
			EDSR-baseline (2017) \cite{lim2017enhanced} & N/A & N/A & 27.57 & 0.7357 & 5.913 & ~~1.018 & 0.015 \\
			CARN (2018) \cite{ahn2018fast} & 27.58 & 0.7349 & 27.58 & 0.7358 & 6.071 & ~~1.006 & 0.017 \\
			FRSR (2019) \cite{soh2019natural} & 27.60 & 0.7366 & 27.61 & 0.7374 & 5.853 & ~~1.284 & 0.028 \\
			EUSR (2018) \cite{kim2018deep} & 27.69 & 0.739~~~ & 27.69 & 0.7403 & 5.964 & ~~1.790 & 0.036 \\
			EDSR (2017) \cite{lim2017enhanced} & 27.71 & 0.7420 & 27.73 & 0.7422 & 5.884 & ~~8.806 & 0.133 \\
			EUSR+ (2018) \cite{kim2018deep} & 27.74 & 0.741~~~ & 27.75 & 0.7415 & 6.019 & 14.223 & 0.249 \\
			RCAN (2018) \cite{zhang2018image} & 27.77 & 0.7436 & 27.75 & 0.7432 & 5.921 & ~~6.607 & 0.102 \\
			EDSR+ (2017) \cite{lim2017enhanced} & 27.79 & 0.7437 & 27.81 & 0.7439 & 5.981 & 68.570 & 0.977 \\
			RCAN+ (2018) \cite{zhang2018image} & 27.85 & 0.7455 & 27.83 & 0.7451 & 6.013 & 52.451 & 0.800 \\
			RRDB (2018) \cite{wang2018esrgan} & 27.85 & 0.7455 & 27.85 & 0.7455 & 5.967 & ~~8.105 & 0.106 \\
			RRDB (2018) \cite{wang2018esrgan} + self-ensemble & N/A & N/A & 27.90 & 0.7466 & 6.008 & 65.631 & 0.794
		\end{tabular}
	\end{center}
	\vspace{-6pt}
	\caption{Performance comparison of selected super-resolution models included in SRZoo, where a scaling factor of 4 is used on the BSD100 dataset \cite{martin2001database}. The models are sorted in terms of the measured PSNR values. The running time is measured with Intel i7-7700K (CPU) and NVIDIA GeForce GTX 1080 (GPU).}
	\label{table:performance_comparison}
\end{table*}

\section{Key Features of SRZoo}
\label{sec:features}

We implement SRZoo by using TensorFlow with Python to achieve the three principal features with ensuring compatibility with various hardware platforms, which are explained in this section.

\subsection{Model conversion}

The main objective of SRZoo is to provide various state-of-the-art deep learning-based super-resolution models in the same repository.
It enables researchers to evaluate the performance of the ``official'' pre-trained models, i.e., the model parameters provided by the original authors, under the same condition.
To do this, SRZoo provides 26 pre-trained super-resolution models\footnote{As of October 2019; more to be added if available.}.
They are initially implemented in different deep learning frameworks, including PyTorch and TensorFlow.
These models are converted for SRZoo using open-sourced model conversion tools, e.g., pytorch2keras \cite{pytorch2keras} to convert PyTorch-based models.
Since these tools are optimized for the image classification models, we modify them to support converting the super-resolution models.
We provide the modified code along with documentation to enable the conversion of the other models.

\subsection{Model configuration}

SRZoo supports various model configurations for the super-resolution tasks.
For instance, SRZoo supports various upscaling factors, e.g., $\times2$, $\times3$, $\times4$, and $\times8$.
In addition, it supports the geometric self-ensemble method, which is widely used for performance boosting during test \cite{lim2017enhanced,zhang2018image,kim2018deep}.
Thus, it is possible to apply the method to the super-resolution models that do not employ it originally.

\subsection{Performance evaluation}

Although most recent super-resolution methods are implemented in deep learning frameworks rather than MATLAB, almost every method is separately evaluated with MATLAB-based codes.
In addition, as we aforementioned in Section~\ref{sec:different_evaluation_criteria}, different evaluation criteria are used to compare the performance directly.
To alleviate this, SRZoo provides Python-based evaluation codes to measure the performance of various super-resolution methods in one place and compare them more equitably than before.
It enables to evaluate the performance within the same environment in terms of both hardware and software configurations.
Besides, it is easy to add a new evaluation metric, which can be done by adding a new ``Evaluator'' class to SRZoo.
Furthermore, SRZoo also supports comparing the models in terms of computation speed, along with the image quality.

\section{Performance comparison}
\label{sec:performance}

\begin{figure}[t]
	\begin{center}
		\centering
		\begin{minipage}[b]{0.495\linewidth}
			\centering
			\centerline{\includegraphics[width=1.0\linewidth]{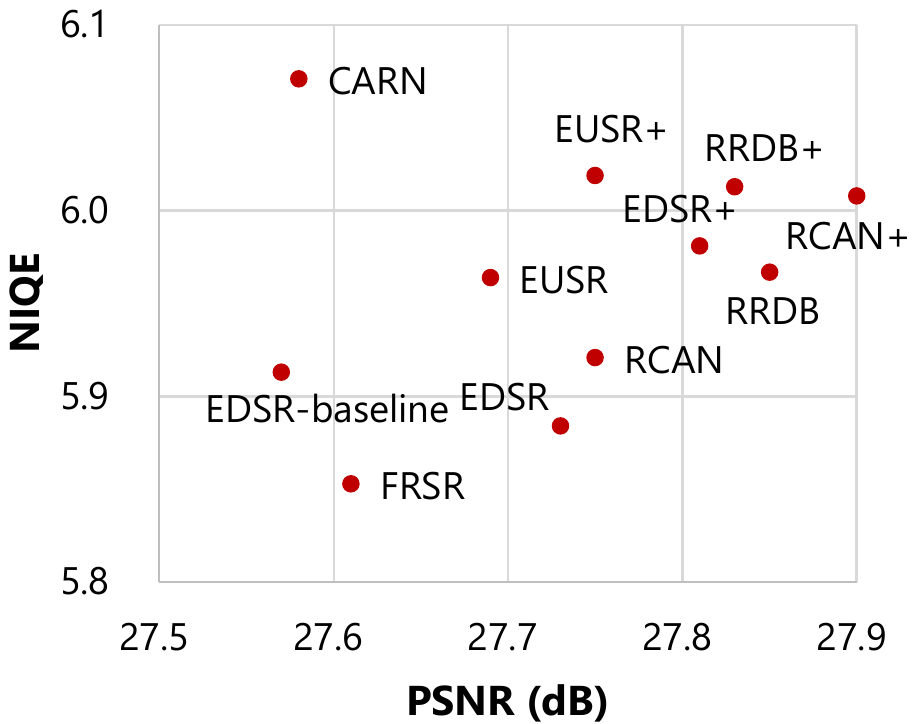}}
			\centerline{(a)}
		\end{minipage}
		\begin{minipage}[b]{0.495\linewidth}
			\centering
			\centerline{\includegraphics[width=1.0\linewidth]{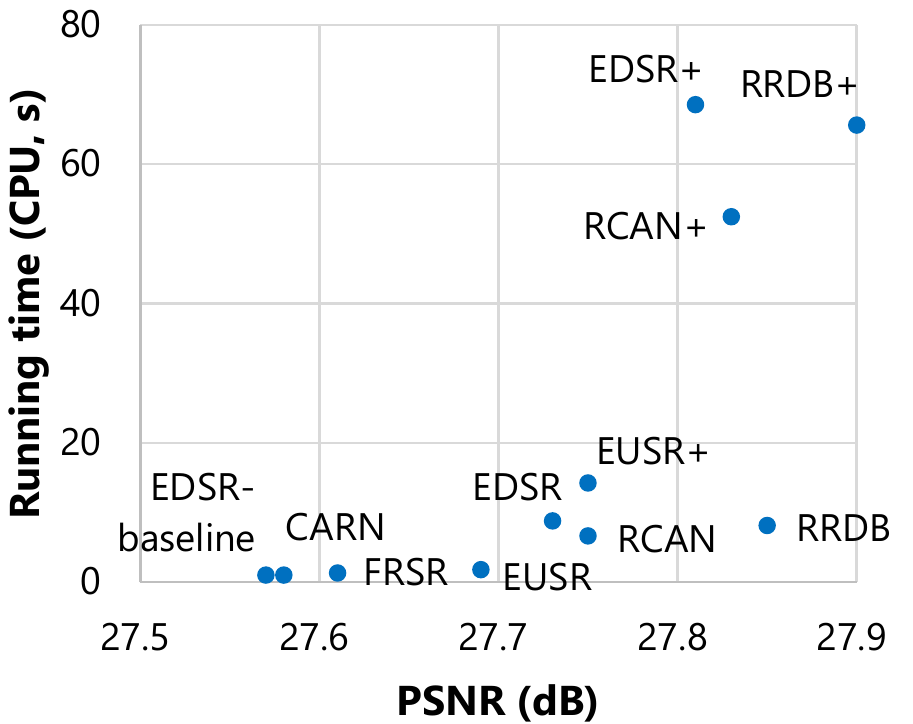}}
			\centerline{(b)}
		\end{minipage}
	\end{center}
	\vspace{-6pt}
	\caption{Performance comparison of super-resolution models using SRZoo. (a) PSNR vs. NIQE (b) PSNR vs. running time on a CPU}
	\label{fig:performance}
\end{figure}

Since the state-of-the-art super-resolution models are included in the single repository, it is possible to make a fair comparison of their performance with SRZoo.
Table~\ref{table:performance_comparison} shows a performance comparison of selected super-resolution models included in SRZoo.
A total of 100 images in BSD100 \cite{martin2001database} are used, which is one of the widely used benchmarking datasets for super-resolution.
The models are run on Intel i7-7700K (CPU) and NVIDIA GeForce GTX 1080 (GPU).
The peak signal-to-noise ratio (PSNR), SSIM, and natural image quality evaluator (NIQE) \cite{mittal2013making} values are measured for the super-resolved outputs.
We follow the most widely used conditions in Table~\ref{table:evaluation_criteria}.
In particular, we calculate the quality metrics on the Y channel of the YCbCr color space with the floating-point precision and shave the image edges with the number of pixels corresponding to the upscaling factor.

We also include the quality metrics that are reported from the original papers of the compared models in Table~\ref{table:evaluation_criteria}.
Note that the PSNR and SSIM values of ESRGAN and EDSR-baseline are not reported in the original papers.
The differences between the reported and measured metrics justify the necessity of SRZoo for a fair comparison of the models with the same criteria.
In addition, SRZoo can evaluate the models with metrics that are not used in the original papers, including NIQE and running time.
Furthermore, SRZoo can boost the performance of the existing super-resolution models.
For instance, RRDB with employing the geometric self-ensemble method shows the best performance in terms of PSNR and SSIM, even though the original implementation does not employ the self-ensemble method.

\figurename~\ref{fig:performance} depicts the performance comparison of the models.
We exclude ESRGAN, whose PSNR value is much lower than those of the others.
Note that a smaller value of NIQE means better performance.
In \figurename~\ref{fig:performance}a, it is observed that there is no significant correlation between the PSNR and NIQE values.
In other words, models having larger PSNR values (i.e., less distortion) are not guaranteed to have smaller NIQE values (i.e., higher perceptual quality), as also noted in \cite{blau2018perception}.
Besides, employing geometric self-ensemble is harmful to perceptual quality, even though it is beneficial to reduce the distortion.
\figurename~\ref{fig:performance}b shows that the models producing images having higher quality tend to have higher computational complexity.
Nevertheless, the RRDB model shows better performance than EDSR+ and RCAN+ in terms of both PSNR and running time.
These results demonstrate that SRZoo enables to analyze the performance of the super-resolution models comprehensively.

\section{Applications}
\label{sec:applications}

\subsection{Platform-agnostic image reconstruction}

\begin{figure}[t]
	\begin{center}
		\centering
		\begin{minipage}[b]{0.72\linewidth}
			\centering
			\centerline{\includegraphics[width=0.99\linewidth]{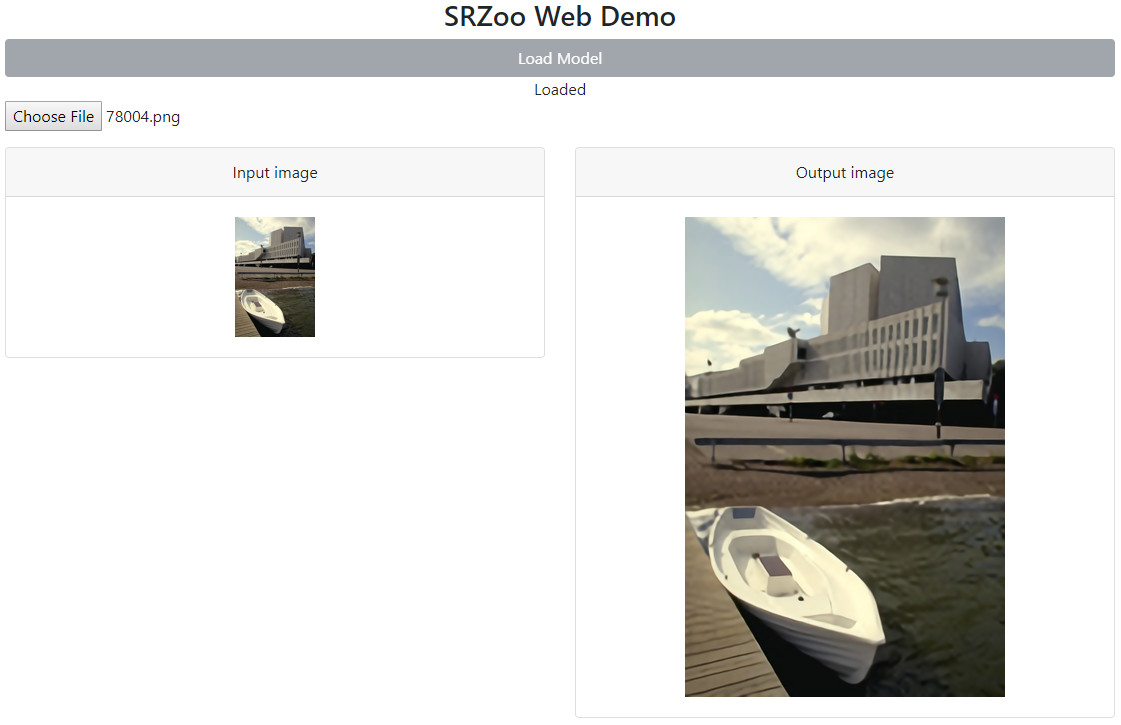}}
			\centerline{(a)}
		\end{minipage}
		\begin{minipage}[b]{0.22\linewidth}
			\centering
			\centerline{\includegraphics[width=0.99\linewidth]{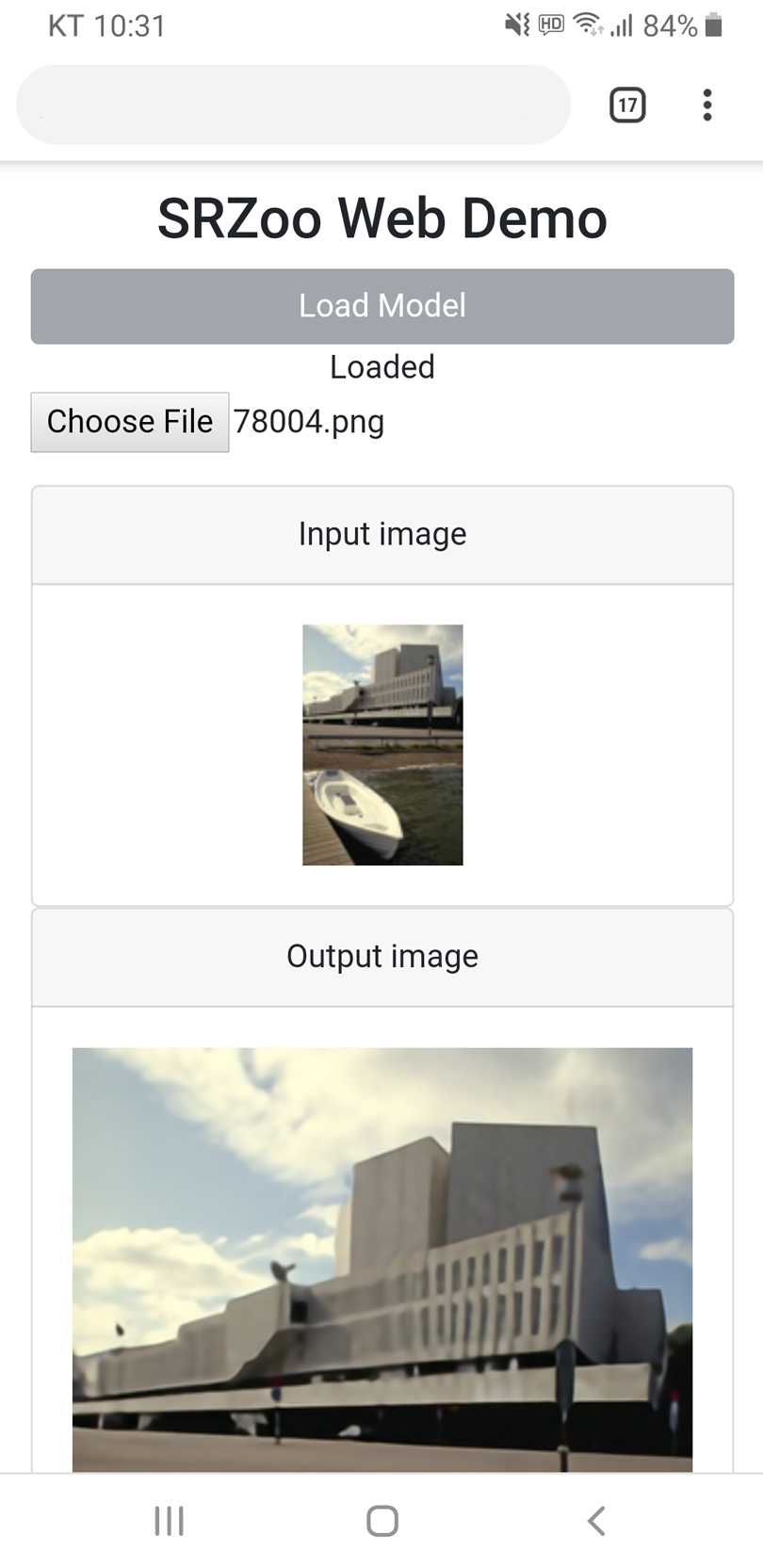}}
			\centerline{(b)}
		\end{minipage}
	\end{center}
	\vspace{-6pt}
	\caption{Image super-resolution using SRZoo running on (a) a desktop web browser and (b) a mobile web browser}
	\label{fig:mobile_application}
\end{figure}

SRZoo provides the pre-trained super-resolution models in a ``ready-for-deployment'' condition.
For various applications of super-resolution, SRZoo allows obtaining super-resolved images regardless of the specific platforms.
For example, the models in SRZoo can be deployed on the mobile platform via TensorFlow Lite or TensorFlow.js.
\figurename~\ref{fig:mobile_application} shows example showcases where the super-resolved output for a given image is produced on both desktop and mobile web browsers.
In addition, the capability of platform-agnostic image reconstruction also allows evaluating the performance of the super-resolution models on various platforms.

\subsection{Extension to advanced topics}

Along with employing the super-resolution models as a standalone application, it is also possible to employ them as a part of other tasks.
For instance, super-resolution can be used as a pre-processing tool to enhance the performance of other image-related tasks, e.g., image classification and image captioning \cite{yin2018deep}.
Since SRZoo provides a separate class for loading the models, it can serve as a ``plugin'' for such tasks without any additional work.
In addition, the provided super-resolution models can be used to thoroughly analyze their intermediate processes by computing the features and gradients.
One example use case is to examine the robustness of the super-resolution models by adding a small perturbation to a given input image, which erroneously deteriorates the super-resolved outputs \cite{choi2019evaluating}.
Therefore, SRZoo can be employed as a testbed to analyze and refine the state-of-the-art deep learning-based super-resolution methods.

\subsection{Extension to other manipulation models}

SRZoo can be easily extended to other image processing tasks thanks to a well-structured architecture.
There are various image manipulation algorithms, including image deblurring, image style transfer, and image compression or decompression.
These take an input image and produce an output image having improved quality or different characteristics.
Since SRZoo is developed to deal with models considering images as both inputs and outputs (while image classification models are not), it can be used to employ such algorithms with only a few modifications.
Similar to the super-resolution models, only a pre-trained model with a configuration file that contains properties of the model are required.
As a proof-of-concept, we provide a pre-trained deep learning-based image compression model \cite{mentzer2018conditional} in SRZoo.
SRZoo considers that the model as an image processing algorithm having an upscaling factor of 1; thus, no modification of the original SRZoo code is necessary.

\section{Conclusion}

In this paper, we proposed SRZoo, which is developed for deep learning-based state-of-the-art super-resolution models.
SRZoo aims to provide an integrated repository that contains various pre-trained super-resolution models, which are ready for deployment, and evaluation tools run in one place.
It enables us to overcome the limitations in employing various super-resolution methods implemented in different deep learning frameworks and to perform a fair comparison of the models using the same evaluation criteria.
We showed the main features of SRZoo in terms of handy model conversion, various model configurations, and fair performance evaluation.
In addition, we suggested possible applications of SRZoo.

\bibliographystyle{IEEEbib}
\bibliography{refs}

\end{document}